\documentclass[10pt,conference]{IEEEtran}
\IEEEoverridecommandlockouts

\usepackage{lipsum}

\usepackage{hyperref}

\usepackage{cite}
\usepackage{color}
\usepackage{amsmath, bm}
\usepackage{amssymb}
\usepackage{amsfonts}
\usepackage{enumitem}
\usepackage[T1]{fontenc}
\usepackage{xspace} 
\usepackage{subfigure}
\usepackage{tcolorbox}
\usepackage{graphicx}
\usepackage{textcomp}
\usepackage{pifont}
\usepackage{booktabs}
\usepackage{multirow}
\usepackage{adjustbox}
\usepackage{makecell}
\usepackage[referable]{threeparttablex}
\usepackage{balance}
\usepackage{tabularx}

\usepackage{stackengine}

\usepackage{listings}
\definecolor{codegreen}{rgb}{0,0.6,0}
\definecolor{codegray}{rgb}{0.5,0.5,0.5}
\definecolor{codered}{rgb}{1,0,0}
\definecolor{codepurple}{rgb}{0.58,0,0.82}
\definecolor{backcolour}{rgb}{1,1,1}

\usepackage{cleveref}

\newcommand{\inlinecode}[1]{\text{\texttt{\small #1}}}

\newcommand{\app}{\textsc{LISP}\xspace}

\newcommand{\ie}{\emph{i.e.,}\xspace}
\newcommand{\eg}{\emph{e.g.,}\xspace}
\newcommand{\etal}{et al.\xspace}

\newcommand{\rev}[1]{#1}

\newcommand{\mut}{API under test\xspace}

\newcommand{\llmsfn}{large language models\xspace}
\newcommand{\LLMfn}{Large Language Model\xspace}
\newcommand{\LLMsfn}{Large Language Models\xspace}

\newcommand{\ecp}{input space partitioning\xspace}
\newcommand{\Ecp}{Input space partitioning\xspace}
\newcommand{\ECP}{Input Space Partitioning\xspace}

\newcommand{\oi}{object instantiation\xspace}

\newcommand{\tdtda}{top-down type dependency analysis\xspace}
\newcommand{\Tdtda}{Top-down type dependency analysis\xspace}
\newcommand{\TdTDA}{Top-down Type Dependency Analysis\xspace}
\newcommand{\buoi}{bottom-up object instantiation\xspace}

\newcommand{\BuOI}{Bottom-up Object Instantiation\xspace}

\newcommand{\cOOPlangs}{common OOP languages\xspace}

\newcommand{\apinum}{2,205\xspace}
\newcommand{\libnum}{10\xspace}
\newcommand{\cvenum}{13\xspace}
\newcommand{\woisp}{w/o ISP\xspace}

\newcommand{\wotdaoi}{w/o TDA+OI\xspace}

\newcommand{\ubs}{exceptions\xspace}
\newcommand{\ub}{exception\xspace}
\newcommand{\ubtype}{18\xspace}
\newcommand{\ubtotal}{1287\xspace}
\newcommand{\ubnum}{404\xspace}

\newcommand{\Impllevel}{Code level\xspace}
\newcommand{\impllevel}{code level\xspace}

\newcommand{\implleveldash}{code-level\xspace}
\newcommand{\Ccptlevel}{Semantic level\xspace}

\newcommand{\ccptleveldash}{semantic-level\xspace}

\def\BibTeX{{\rm B\kern-.05em{\sc i\kern-.025em b}\kern-.08em
    T\kern-.1667em\lower.7ex\hbox{E}\kern-.125emX}}

\begin{document}
\title{LLM Based Input Space Partitioning Testing for Library APIs}

\author{\IEEEauthorblockN{Jiageng Li}
\IEEEauthorblockA{\textit{Fudan University} \\
jgli22@m.fudan.edu.cn}
\and
\IEEEauthorblockN{Zhen Dong\IEEEauthorrefmark{1}\thanks{\IEEEauthorrefmark{1} Corresponding author.}}
\IEEEauthorblockA{\textit{Fudan University} \\
zhendong@fudan.edu.cn}
\and
\IEEEauthorblockN{Chong Wang}
\IEEEauthorblockA{\textit{Nanyang Technological University} \\
chong.wang@ntu.edu.sg 
}
\and
\IEEEauthorblockN{Haozhen You}
\IEEEauthorblockA{\textit{Fudan University} \\
hzyou23@m.fudan.edu.cn}
\and
\IEEEauthorblockN{Cen Zhang}
\IEEEauthorblockA{\textit{Nanyang Technological University} \\
cen001@e.ntu.edu.sg}
\and
\IEEEauthorblockN{Yang Liu}
\IEEEauthorblockA{\textit{Nanyang Technological University} \\
yangliu@ntu.edu.sg}
\and
\IEEEauthorblockN{Xin Peng}
\IEEEauthorblockA{\textit{Fudan University} \\
pengxin@fudan.edu.cn}
}

\maketitle

\begin{abstract}

Automated library APIs testing is difficult as it requires exploring a vast space of parameter inputs that may involve objects with complex data types. Existing search based approaches, with limited knowledge of relations between object states and program branches, often suffer from the low efficiency issue, \ie tending to generate invalid inputs. Symbolic execution based approaches can effectively identify such relations, but fail to scale to large programs.

In this work, we present an LLM-based input space partitioning testing approach, \app, for library APIs. The approach leverages LLMs to understand the code of a library API under test and perform input space partitioning based on its understanding and rich common knowledge. Specifically, we provide the signature and code of the \mut to LLMs, with the expectation of obtaining a text description of each input space partition of the \mut. Then, we generate inputs through employing the generated text description to sample inputs from each partition, ultimately resulting in test suites that systematically explore the program behavior of the API.

We evaluate \app on more than \apinum library API methods taken from 10 popular open-source Java libraries (\eg apache/commons-lang with 2.6k stars, guava with 48.8k stars on GitHub). Our experiment results show that \app is effective in library API testing. It significantly outperforms state-of-the-art tool EvoSuite in terms of edge coverage.
On average, \app achieves 67.82\% branch coverage, surpassing EvoSuite by 1.21 times.
In total, \app triggers 404 exceptions or errors in the experiments, and discovers \cvenum previously unknown vulnerabilities during evaluation, which have been assigned CVE IDs.

\end{abstract}


\maketitle

\begin{IEEEkeywords}
Input Space Partitioning Testing, Large Language Models, Symbolic Execution, API testing.
\end{IEEEkeywords}

\section{Introduction~\label{sec:intro}}

The third party libraries, as an essential part in software ecosystems, have become one of the most significant contributors to fast development of today's software system. According to a recent study~\cite{wen}, a Java project directly relies on different 14 third party libraries. Vulnerabilities within these libraries can pose significant risks to numerous software systems. Consequently, testing libraries is imperative to ensure system security. 

However, testing library APIs is notoriously challenging as it entails exploring a vast input space of multiple parameters, particularly when these parameters involve objects with complex data types. The behavior of the libraries can be constrained by a specific state of one or more input objects. Triggering such a state involves generating input values satisfying \emph{relevant} conditions as well as generating statements to instantiate these objects. 

This poses numerous challenges for existing automated test generation techniques: (1) \emph{Search based testing:} Most existing techniques~\cite{ltlfuzz,2011evosuite,timemachine,symbolicdroid,setdroid,detectdroid} frame automated test generation as an optimization problem over the input space with the goal of generating inputs to achieve maximal code coverage, for instance, EvoSuite~\cite{2011evosuite}, a widely used automated test generation tool adopts a genetic algorithm to generate tests. The problem with this type of techniques is the low efficiency issue when tackling the expansive space of inputs involving multiple objects with complex data types within library APIs; (2) \emph{Symbolic execution:} Symbolic execution is an effective testing technique that can generate inputs that cover desired program paths. 
Yannic Noller \etal leverages symbolic execution to guide fuzzing to generate inputs that cover deep program behavior~\cite{noller2018badger}. Despite significant efficiency improvement, these techniques face difficulties in scaling to large programs due to inherent limitations of symbolic execution,
\rev{\eg SPF~\cite{2019svcompspf}  has limited support for heap input.}
SUSHI~\cite{2018icsesushi} proposed by Pietro Braione \etal can be only applied to Java classes.

In this paper, we view automated test generation as a program input space sampling problem. The ideal way to sample is to compute \emph{input space partitions}, and then choose inputs from each partition so as to cover all possible program behavior. In this perspective, search-based approaches kind of leverage heuristics to guide search, aiming to sample inputs from as many partitions as possible. Symbolic execution based approaches attempt to compute input space partitions by solving program path conditions and then sample inputs from each partition. Both type of approaches come at a cost. The former requires executing a large amount of inputs that go through redundant program paths, the latter requires heavy computation resources to solve path conditions. 

In this work, we propose an \LLMfn (LLM) based input space partitioning testing approach for library APIs. Specifically, we leverage LLMs to infer the input space partitions of a library API under test and then sample inputs from each partition so as to generate test suites with high quality. Recently, LLMs have demonstrated promising capabilities in understanding programs and common knowledge reasoning, leading to their widespread adoption in the software engineering domain~\cite{yuan2023evaluating, lemieux2023codamosa}. \emph{Motivated by these capabilities, we explore using LLMs to automate input space partitioning, achieving the objectives of symbolic execution without explicitly performing it}. To this end, we propose a framework that interacts with LLMs to compute input space partitions for a given library API and generate input values based on textual descriptions of each partition, resulting in high-quality test inputs. Subsequently, the framework takes those inputs to generate test suites for library API testing. 

We evaluated \app on \apinum  APIs from 10 widely used libraries, including Apache Commons-lang3 and Google Guava. The results show \app is highly effective in testing library APIs, achieving exceptionally high code coverage with a minimal number of generated inputs.  In the comparison experiments, \app outperformed the state-of-the-art technique EvoSuite, achieving 1.21 times higher edge coverage. Furthermore, \app identified 404  exceptions across the 10 libraries, including \cvenum previously undiscovered vulnerabilities, which have been assigned CVE IDs. To support future research, we make our the experimental data and results publicly available at the following link: \url{https://github.com/FudanSELab/LISP}~\cite{site}.

\section{Motivating Examples}\label{sec:motivation}

\begin{lstlisting}[caption={org.apfloat.ApcomplexMath::pow}, label={lst:pow}]
// ApcomplexMath.java
public static Apcomplex pow(Apcomplex z,Apcomplex w)
throws ApfloatRuntimeException {
  Apcomplex result = ApfloatHelper.checkPow(
    z, w, Math.min(z.precision(), w.precision()));
  if (result != null) {
    return result;
  } else if (z.real().signum() >= 0 &&
             z.imag().signum() == 0) {
    Apfloat x = z.real();
    Apfloat one = new Apfloat(
      1L, Long.MAX_VALUE, x.radix());
    x = // ignore some code
    return exp(w.multiply(ApfloatMath.log(x)));
  } else {
    return exp(w.multiply(log(z)));
  }
}
\end{lstlisting}

\subsection{Importance of Code Understanding and Common Knowledge}

\Cref{lst:pow} presents an API method named \inlinecode{pow} within the \inlinecode{class ApcomplexMath} from the \textit{apfloat}. This method, which takes two parameters named \inlinecode{z} and \inlinecode{w}, exhibits distinct behaviors based on the content of \inlinecode{z} and \inlinecode{w}, which means that each input space can be represented by the states of \inlinecode{z} and \inlinecode{w}. Specifically, when \inlinecode{result != null}, the API returns the \inlinecode{result} directly (line 7); when \inlinecode{z.real().signum()} is greater than or equal to $0$ and \inlinecode{z.imag().signum()} equals $0$ (line 8-9), the API returns the result at line 14. Otherwise, the API engages in a calculation for complex numbers (line 16).

In software testing, precise partitioning of the input space facilitates efficient input generation.
\begin{itemize}[leftmargin=10pt]
    \item \textit{Symbolic execution} is the ideal solution to partition the input space. We attempt one of the state-of-the-art tools, SPF. However, it fails to work due to insufficient modeling of native methods when creating an \inlinecode{Apcomplex} object.
    \item \textit{Search-based testing} is another approach for \ecp, which is more scalable compared to symbolic execution. \rev{We use the state-of-the-art tool in the SBST field, EvoSuite, with the default configuration and run it for 200s. EvoSuite generates 64 test cases but only achieves 55\% coverage. We find that EvoSuite generates a large number of equivalent inputs, none of which can reach line 10-14.
\end{itemize}

In the context of ``exponentiation'', awareness of certain corner cases is crucial. For instance, $0^0$ is typically undefined; \rev{the computation process of $z^w (z \in R)$ is different from that of $z^w (z \not\in R$)}. Failure to bridge the gap between such background knowledge and software testing, leads to blind exploration of a vast search space for \inlinecode{z} and \inlinecode{w}.
Therefore, it is essential to present an approach that effectively understands and navigates the input space while avoiding falling into the trap of generating invalid or single-scenario inputs.
This approach should integrate common knowledge in both \implleveldash and \ccptleveldash.
}

\subsection{\ECP with \LLMsfn}
Recently, \llmsfn (LLMs) have demonstrated considerable capabilities across diverse domains, such as code understanding~\cite{lingmingissta2022understanding, niu2023empirical}, common knowledge acquisition~\cite{Radford2019LanguageMA, chowdhery2022palm, gao2020pile, austin2021program, nijkamp2023codegen} and code generation~\cite{deng2023large, deng2023fuzzgpt}, which align with the requirements for library API testing.

Assume that we need to test the \inlinecode{pow} method. We can employ LLMs to partition the input space. Specifically, we can provide the signature and code of \inlinecode{pow} for LLMs and instruct them to partition the input space. Then, we can obtain the text form of the input space partitioning results, such as ``
(1) \inlinecode{z}: real part is non-negative and imaginary part is 0; \inlinecode{w}: is an Apcomplex number.
(2) \inlinecode{z}: real part is negative or imaginary part is non-zero; \inlinecode{w}: is an Apcomplex number''. From the above input space partitioning results, we know that LLMs believes that it should generate a complex number with ``Real positive and Imaginary zero'', which is exactly one of the conditions for entering a block (line 10-14) in \Cref{lst:pow} (another implicit condition is ``\inlinecode{result == null}'').

\subsection{Input Generation with \LLMsfn}

\Cref{lst:apfloat-cons} presents two types, \inlinecode{Apcomplex} and \inlinecode{Apfloat}. \inlinecode{Apcomplex} inherits \inlinecode{java.lang.Number} and represents complex numbers~\cite{enwiki:1187479457complexnumber} in mathematics. \inlinecode{Apfloat} inherits the former type and represents float numbers. In addition, we present two constructors of type \inlinecode{Apcomplex}, and the first constructor requires inputs of type \inlinecode{Apfloat}.

Assume that we intend to construct a corresponding \inlinecode{Apcomplex} instance for the parameter \inlinecode{z}, which complies with the requirements of the text description of this input space partition (``real part is non-negative and imaginary part is 0''). In general, the process can be divided into two necessary steps.

\begin{lstlisting}[caption={Type \texttt{Apcomplex} and Type \texttt{Apfloat}}, label={lst:apfloat-cons}]
public class Apcomplex extends Number {
    private Apfloat real;
    private Apfloat imag;
    public Apcomplex(Apfloat real, Apfloat imag) ..
    public Apcomplex(String value) ..
    // overlook other constructors
}

public class Apfloat extends Apcomplex {
    private ApfloatImpl impl;
    public Apfloat(long value) ..
    public Apfloat(String value, long precision) ..
    // overlook other constructors
}
\end{lstlisting}

\subsubsection{Top-down type dependency analysis and constructor selection}

\rev{To generate inputs for a reference type, we need to acquire 
\rev{(1) all derived classes of that type, and (2) all constructors of any involved reference types.}
For this example, to generate an input object of \inlinecode{Apcomplex} type representing $1 + 0i$, we first retrieve its available constructors and then identify the appropriate constructors. This process continues recursively until all relevant reference types are addressed,
\rev{resulting in a sequence of constructors that can be used to generate the target object.}
Specifically, we provide LLMs with the text description of partition, so as to drive LLMs to select the appropriate constructors. In \rev{\Cref{lst:apfloat-cons}}, the first constructor that takes \inlinecode{real} and \inlinecode{imag} as two parameters, is exactly what we need. Since the type of \inlinecode{real} and \inlinecode{imag} is still a reference type, we repeat the previous process to generate two \inlinecode{Apfloat} objects. In this case, we use LLMs to select three constructors, as depicted in the upper half of \rev{\Cref{lst:tda-oi}}.}

\begin{lstlisting}[caption={Selected Constructors and Instantiation Statements}, label={lst:tda-oi}]
    // selected constructors
    // after type dependency analysis
    Apfloat real = new Apfloat(/* TODO */);
    Apfloat imag = new Apfloat(/* TODO */);
    Apfloat c1 = new Apcomplex(real, imag);
    
    // object instantiation statements
    float real_value = 1.0f;
    float imag_value = 0.0f;
    Apfloat real = new Apfloat(real_value);
    Apfloat imag = new Apfloat(imag_value);
    Apfloat c1 = new Apcomplex(real, imag);
\end{lstlisting}

\subsubsection{Bottom-up \oi with concrete values}

\rev{After obtaining the appropriate constructors, \rev{we need to fill in correct values to generate the desired input object.}
For this example, to instantiate an \inlinecode{Apcomplex} instance representing $1 + 0i$, it is required to construct an \inlinecode{Apfloat} object representing $1$ and another \inlinecode{Apfloat} object representing $0$, according to the selected constructors in the upper half of \rev{\Cref{lst:tda-oi}}.
Specifically, we can provide LLMs with these \rev{selected constructors}, supplemented by a text description of the partition with specific values, so as to guide LLMs to generate valid inputs, as shown in the lower half of \rev{\Cref{lst:tda-oi}}.}

Looking at the process of constructing the \inlinecode{z} for \inlinecode{pow}, LLMs can serve as a vital tool in the field of input space partitioning testing.
Specifically, we have utilized the code understanding and generation capabilities of LLMs in three place, \ie \ecp, \tdtda and \buoi.

\begin{figure*}[htb]
  \centering
  \includegraphics[width=\textwidth]{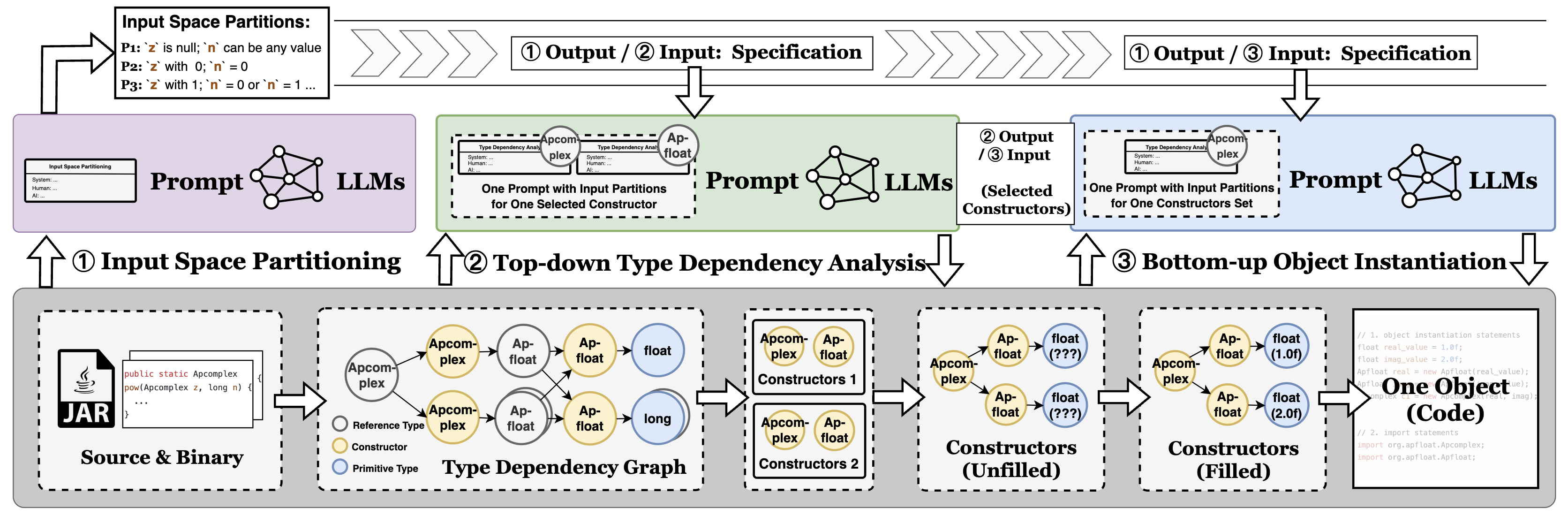}
  \caption{Approach Overview of \app}
  \label{fig:approach}
\end{figure*}

\section{Approach: \app}
\label{sec:approach}

\rev{We introduce \rev{\app}, a novel workflow designed to strategically guide \LLMsfn (LLMs) in understanding the source code of API methods. This approach ultimately generates high-quality inputs and tests \rev{drivers} for library APIs.}

In \Cref{fig:approach}, the lower section (\ie the gray part) illustrates the common process of existing input generation approaches.
We find that search-based approaches typically search and partition the input space at runtime, often neglecting the source code of the APIs~\cite{2018icsesushi}.
Building on these insights, the upper section of \Cref{fig:approach} depicts our proposed workflow, which decomposes the API input generation process into three parts: \textit{\ecp}, \textit{\tdtda}, and \textit{\buoi}.

\subsection{\ECP}

\begin{figure}[htb]
    \centering
    \includegraphics[width=1.0\columnwidth]{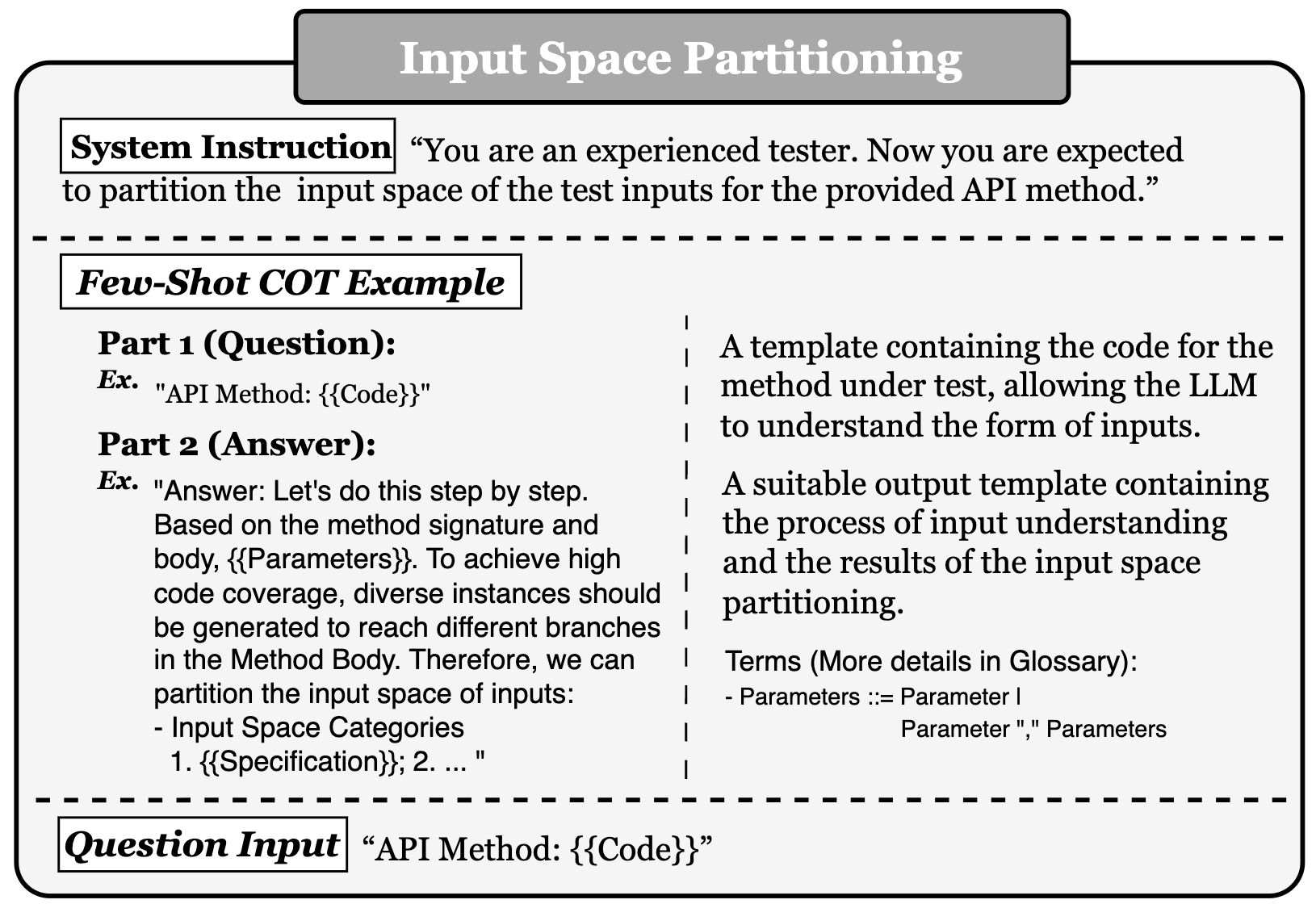}
    \caption{The prompt for \ecp}
    \label{fig:prompt-ecp}
\end{figure}

To systematically generate inputs for a given \mut with the goal of covering all branches and triggering exceptional behaviors efficiently, a thorough understanding of the input search space is crucial.

\begin{itemize}[leftmargin=10pt]
    \item \textit{\Ccptlevel}. Inputs often embody concepts within specific domains (\eg exponentiation), reflecting background knowledge.~\cite{petroni2019language, huang2022learnknowledge} This \ccptleveldash understanding imposes constraints on input values, effectively narrowing and categorizing the input space into distinct partitions. For example, for the parameter \inlinecode{z} of the \inlinecode{pow} function in \Cref{lst:pow}, a valid value should contain two numbers representing the real part and the imaginary part, respectively.
    \item \textit{\Impllevel}. When implementing a functionality, the specific implementation is contingent on factors such as project-specific logic, code optimization strategies, and others. Consequently, the input space is further restricted and partitioned at \impllevel. For example, \inlinecode{pow} incorporates a special branch for the parameter \inlinecode{z} to handle the case where \inlinecode{z} represents a positive real number.
\end{itemize}

\rev{We have designed a prompt to harness the capacities of LLMs at the semantic and code levels.
The prompt is illustrated in \Cref{fig:prompt-ecp}.}

\begin{itemize}[leftmargin=10pt]
    \item \textit{System Instruction}. We highlight ``\ecp'' as the task we expect the LLM to accomplish.
    \item \textit{Few-shot CoT Examples}. 
    \begin{itemize}[leftmargin=10pt]
        \item \textit{Question}. We include only the source code of the \mut.
        We emphasize that the source of the \mut encapsulates both \ccptleveldash and \implleveldash knowledge to understand the input space of its parameters.
        In addition, we also implement a \app variant, described in \Cref{sec:eval-setup}, which includes the called methods based on the call graph, in order to provide more contexts. This variant is called \app-CG.
        \item \textit{Answer}. We construct a chain of thought that analyzes the code in the order of method signature, body, and parameters. We expect the LLM to understand the code under our guidance and provide partitioning results for achieving high coverage.
    \end{itemize}
    
\end{itemize}

\rev{For this part, the input is the source \textit{Code} (\Cref{tab:keywords}) of the \mut, and the output is a collection of textual descriptions of the input space partitions, referred to as \textit{Specification}s  (\Cref{tab:keywords}).}

\rev{\textit{Example.}
For the \inlinecode{pow} API presented in \Cref{lst:pow}, \app can produce 6 partitions of the input space.
These partitions are represented in textual form, \eg ``z: real part is non-negative and imaginary part is 0; w: is an Apcomplex number'', which covered line 10-14.}

\begin{table*}[htb]
\caption{Glossary of Keywords in Prompts}
\centering
\begin{tabularx}{\textwidth}{ccll}
\toprule
No & Keyword & Description & Example \\
\midrule
1 & Code & \rev{The source code of the \mut.} & \rev{``public static Apfloat pow(Apcomplex z, Apcomplex w) \{ ... \}''} \\
2 & Type & \rev{The fully-qualified name of a type.} & \rev{``org.apfloat.Apfloat'', ``org.apfloat.Apcomplex''} \\
3 & Parameter & \rev{The parameter list of the API method under test.} & \rev{``Apcomplex z'', ``Apcomplex w''} \\
4 & Constructor & The constructor of a \textit{Type}. & \rev{``Apcomplex(Apfloat real, Apfloat imag)''} \\
5 & Dependency &
\begin{tabular}[c]{@{}l@{}}
    The ``is-a'' and ``has-a'' relationships \\
    between two \textit{Type}s. \rev{(represented as text)}
\end{tabular} &
\begin{tabular}[c]{@{}l@{}}
    ``class org.apfloat.Apfloat: Constructors: \\
    \quad public Apfloat(float value)''
\end{tabular}
\\
6 & Specification & The constraints on the input space partition. & ``z with 1; n = 0 or n = 1'' \\
\bottomrule
\end{tabularx}
\label{tab:keywords}
\end{table*}

\subsection{\TdTDA}

\begin{figure}[htb]
    \centering
    \includegraphics[width=1.0\columnwidth]{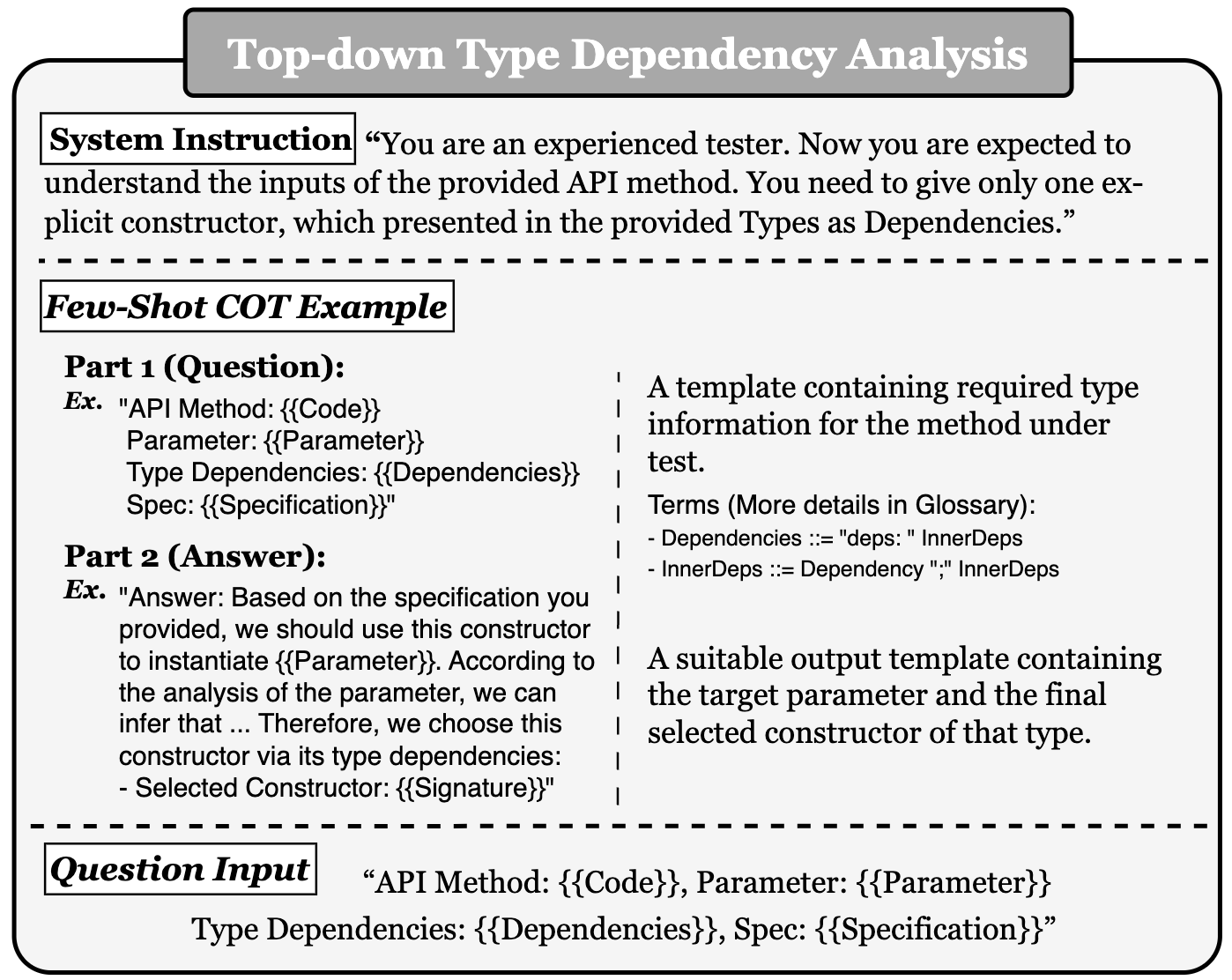}
    \caption{The prompt for \tdtda}
    \label{fig:prompt-cc}
\end{figure}

The type of each parameter in the \mut typically can be classified into the primitive type and the reference type. For the primitive type, we can create them directly with language-specific syntax. However, for the reference type, creating an object is not trivial, and the following two categories of issues arise simultaneously.
(1) \textit{Nested Reference Types}. In various \cOOPlangs, the reference type often involve multiple levels of nesting\rev{, which means that constructing an object of a reference type may require multiple calls to constructors.}
(2) \textit{Multiple Constructor Candidates}. Since a type tends to own multiple constructors, different constructors often yield different construction results.

For the first issue, we construct a ``\textit{Type Dependency Graph}'' (TDG). In detail, we abstract each reference type as a ``node'' and view the usage of each reference type during the instantiation of an object as an ``edge'', and select all reachable types derived from the types of parameters in the directed acyclic graph.

For the second issue, we engage in an interaction with the LLM to select the most appropriate constructor, based on the text description of the input space partition (\ie \textit{Specification}).
We design a prompt to drive LLMs to select the most appropriate constructor for each type, along the top-down process. The prompt is illustrated in \Cref{fig:prompt-cc}.

\begin{itemize}[leftmargin=10pt]
    \item \textit{System Instruction}. We highlight ``constructor selection'' as the task and expect that the LLM can employ the \textit{Specification} to select only one \textit{Constructor} (\Cref{tab:keywords}) for each type.

    \item \textit{Few-shot COT Examples}.
    \begin{itemize}[leftmargin=10pt]
        \item \textit{Question}. For each parameter in the \mut, we provide a list of \textit{Constructor}s for each type and attach the corresponding \textit{Specification} along with the dependency information of the parameter type recorded in the TDG.
        \item \textit{Answer}. We expects the LLM not only to take all provided information into consideration, but also to select only one constructor for each type.
    \end{itemize}

\end{itemize}

\rev{For this part, the inputs are the \textit{Code} of the \mut, and the \textit{Specification}s, while the output is a mappings between \textit{Parameter} (\Cref{tab:keywords}) and its corresponding \textit{Constructor} (\Cref{tab:keywords}) sequence used for instantiation.}

\textit{Examples}. For the constructors presented in \rev{\Cref{lst:apfloat-cons}}, \app can output the selected constructors like the upper half of \rev{\Cref{lst:tda-oi}}. Specifically, \app first selects the first constructor for the \inlinecode{Apcomplex} type parameter in the \mut, and then selects the first constructor of \inlinecode{Apfloat} for both ``\inlinecode{Apfloat real}'' and ``\inlinecode{Apfloat imag}'', according to one of partitions outputted by \ecp.
\rev{We break down the type dependency analysis task through the TDG into multiple sub-tasks, which increases the number of interactions with the LLM, but reduces the token of a single prompt, which avoids exceeding the token limit and also helps the LLM focus on selecting the appropriate constructor for a single type.}

\subsection{\BuOI}

\begin{figure}[htb]
    \centering
    \includegraphics[width=1.0\columnwidth]{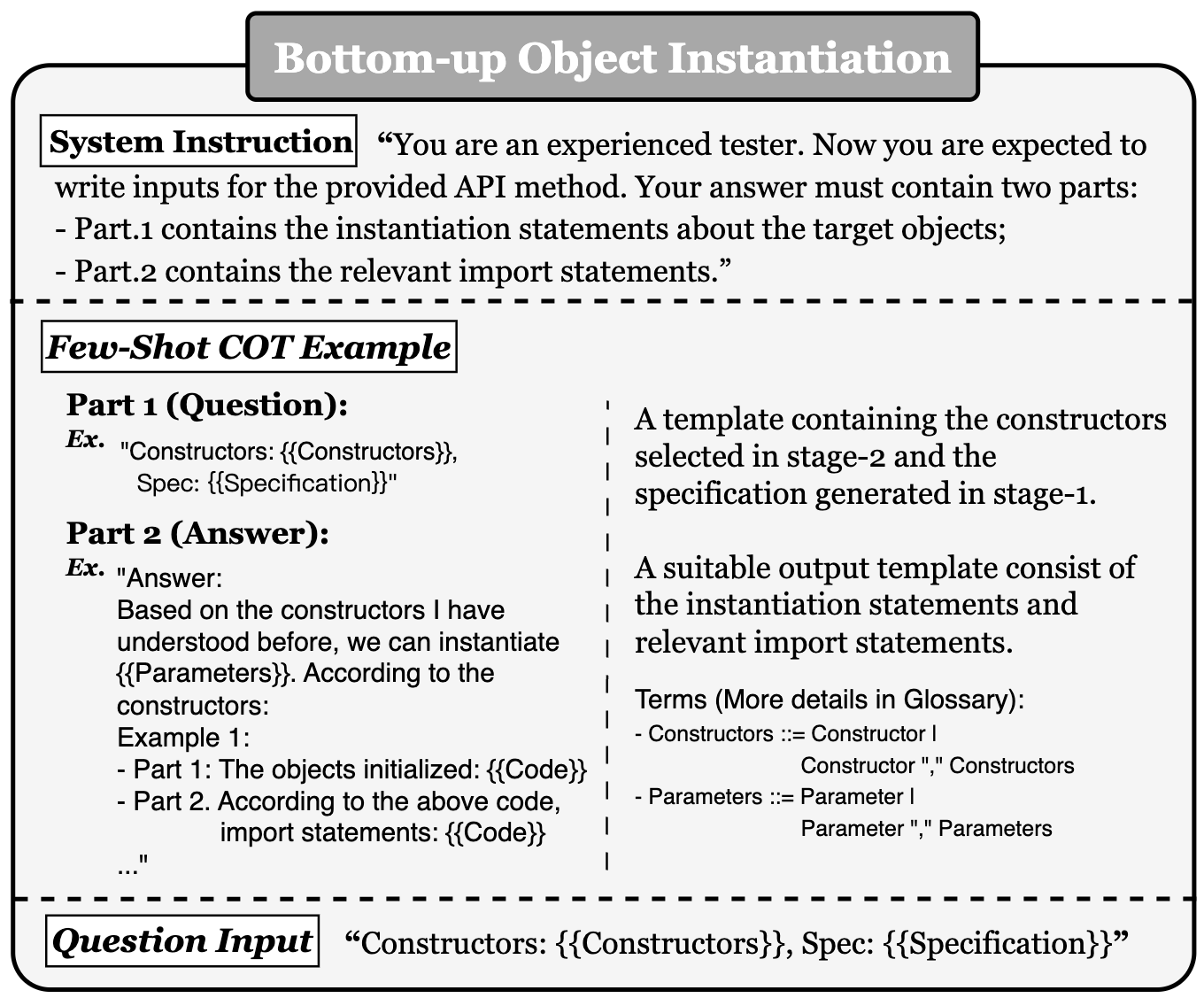}
    \caption{The prompt for \buoi}
    \label{fig:prompt-og}
\end{figure}

The ultimate goal of \app is to generate high-quality inputs. We need to fill in appropriate values into the selected constructors and obtain instantiation statements through interaction with LLMs.
We design a prompt to drive LLMs to fill the appropriate values into the selected constructor. The prompt is illustrated in \Cref{fig:prompt-og}.

\begin{itemize}[leftmargin=10pt]
    \item \textit{System Instruction}. We highlight two parts of statements that the LLM is supposed to provide (1) the ``instantiation statements'' about the target inputs, and (2) the ``import statements'' related to instantiation. 

    \item \textit{Few-shot COT Examples}.
    \begin{itemize}[leftmargin=10pt]
        \item \textit{Question}. We provide all selected \textit{Constructor}s and the \textit{Specification} to assist the LLM in filling in the appropriate values. Then, we consider the objects instantiated in this way as arguments.
        \item \textit{Answer}. \rev{We construct a chain of thought and expect the LLM to synthesize the instantiate statements and the relevant import statements.
        }
    \end{itemize}

\end{itemize}

For this part, the inputs are \textit{Specification}s and selected \textit{Constructor}s, while the outputs are statements that can be used in object instantiation.

\rev{\textit{Examples}. For the selected constructors presented in the upper half of \rev{\Cref{lst:tda-oi}}, if the specification represents \rev{$1 + 0i$, \app can fill ``\inlinecode{1.0f}'' and ``\inlinecode{0.0f}''} into selected constructors and finally generate instantiation statements like in the lower half of \rev{\Cref{lst:tda-oi}}.}

\rev{\textit{Test Driver Generation}. After interacting with the LLM and extracting the statements for constructor invocation, \rev{we encapsulates these statements with the necessary class and method declarations (\ie \inlinecode{class Driver} and \inlinecode{void main(...)}). The generated driver is expected to instantiate objects and invoke the \mut. This process results in the creation of an executable program. The driver template is available~\cite{site}}.}

\section{Evaluation}
\label{sec:evaluation}

We conduct extensive experiments to evaluate \app with the following research questions.

\begin{itemize}[leftmargin=10pt]
    \item \textbf{RQ1 (Code Coverage)}. To what extent can \app cover the code? Can \app outperform the state-of-the-art test tools, EvoSuite and SPF, in terms of code coverage?
    \item \textbf{RQ2 (Usefulness)}. Can \app trigger \ubs? (We only focus on unhandled exceptions and errors, both of which implement \inlinecode{java.lang.Throwable} but used in different scenarios) Can \app find vulnerabilities previously not discovered?
    \item \rev{\textbf{RQ3 (Cost)}. Can \app outperform EvoSuite in terms of time, while keeping the token consumption within a reasonable range?}
    \item \textbf{RQ4 (Ablation Study)}. Are \ecp and \tdtda of \app both effective? How do they contribute?
\end{itemize}

\subsection{Evaluation Setup}\label{sec:eval-setup}

\textit{Experiment Subjects}. To evaluate \app, we selected \libnum Java libraries from previous studies~\cite{2023callmemaybe, xie2023chatunitest} and some awesome lists (\ie \href{https://github.com/akullpp/awesome-java}{awesome-java}, \href{https://github.com/Vedenin/useful-java-links}{useful-java-links}), with the requirement that each selected library is highly starred and has recent code commits. All experimental data and results are available on our site~\cite{site}.

\begin{table}[htb]
\caption{Details of \libnum Java libraries selected. \\
\#$LOC$ = the number of line of code of the library; \\
\#$Stars$ = the number of stars of the \rev{GitHub} repository; \\
\#$APIs$ = the number of selected api methods;
}
\centering
\begin{tabular}{clllll}
\toprule
\#$No$ & Library Name & Version & \#$LOC$ & \#$Stars$ & \#$APIs$ \\
\midrule
1 & commons-lang3 & 3.13.0 & 85.6k & 2.6k & 545 \\
2 & guava & 32.1.2-jre & 163.7k & 48.8k & 195 \\
3 & jfreechart & 1.5.4 & 214.1k & 1.1k & 169 \\
4 & jgrapht & 1.5.2 & 89.8k & 2.5k & 131 \\
5 & joda-time & 2.12.5 & 72.2k & 4.9k & 185 \\
6 & threeten & 1.6.8 & 51.9k & 546 & 106 \\
7 & time4j-base & 5.9.1 & 74.3k & 407 & 70 \\
8 & iCal4j & 4.0.0-rc3 & 66.1k & 705 & 201 \\
9 & SIS-Utility & 1.4 & 783.8k & 94 & 505 \\
10 & XChart & 3.8.7 & 36.9k & 1.5k & 98 \\
\bottomrule
\end{tabular}
\label{tab:projects}
\end{table}

\begin{table*}[htb]
\caption{\rev{Details of Results in RQ1.}}
\centering
\fontsize{7}{10}\selectfont
\begin{tabular}{c|c|cccccccccc|c}
\toprule
\multirow{2}{*}{Metrics} & \multirow{2}{*}{Indicators} & \multicolumn{10}{c|}{Libraries} & \multirow{2}{*}{Overall} \\
\cline{3-12}
 &  & commons-lang3 & JFreeChart & JGraphT & guava & joda-time & threeten & time4j & iCal4j & SIS-Utility & XChart &  \\
\midrule
\multirow{6}{*}{\#$Input$}
 & \app & \rev{3,409} & \rev{694} & \rev{772} & \rev{1,191} & \rev{853} & \rev{505} & \rev{399} & \rev{1,033} & \rev{2,814} & \rev{503} & \rev{12,173} \\
 & EvoSuite-100s & 9,925 & 2,902 & 2,342 & 3,596 & 3,068 & 1,776 & 1,397 & 3,723 & 8,813 & 1,767 & 39,309 \\
 & EvoSuite-150s & 14,343 & 4,155 & 3,226 & 5,164 & 4,645 & 2,743 & 1,756 & 5,209 & 13,031 & 2,545 & 56,817 \\
 & EvoSuite-200s & 18,628 & 5,326 & 4,392 & 6,597 & 6,173 & 3,487 & 2,288 & 6,691 & 16,919 & 3,289 & 73,790 \\
 & LLM-baseline & \rev{1,328} & \rev{162} & \rev{123} & \rev{474} & \rev{301} & \rev{190} & \rev{67} & \rev{279} & \rev{1087} & \rev{130} & \rev{4,141} \\
 & \rev{\app-CG} & \rev{5,104} & \rev{663} & \rev{890} & \rev{1,811} & \rev{906} & \rev{624} & \rev{342} & \rev{1,846} & \rev{3357} & \rev{618} & \rev{16,161} \\
\hline
\multirow{6}{*}{\#$Edge$}
 & \rev{\app} & \rev{7,443} & \rev{1,135} & \rev{1,959} & \rev{1,702} & \rev{1,407} & \rev{586} & \rev{647} & \rev{757} & \rev{3,227} & \rev{237} & \rev{19,100} \\
 & EvoSuite-100s & 5,678 & 1,041 & 1,569 & 1,490 & 476 & 471 & 374 & 673 & 3,278 & 183 & 15,233 \\
 & EvoSuite-150s & 5,724 & 1,065 & 1,754 & 1,571 & 460 & 481 & 353 & 626 & 3,358 & 179 & 15,571 \\
 & EvoSuite-200s & 5,734 & 1,120 & 1,838 & 1,657 & 527 & 518 & 353 & 532 & 3,252 & 194 & 15,725 \\
 & LLM-baseline & \rev{4,255} & \rev{386} & \rev{570} & \rev{1,041} & \rev{829} & \rev{339} & \rev{213} & \rev{417} & \rev{2,336} & \rev{93} & \rev{10,489} \\
 & \rev{\app-CG} & \rev{7,977} & \rev{1,235} & \rev{1,864} & \rev{1,734} & \rev{1,350} & \rev{655} & \rev{749} & \rev{756} & \rev{3,224} & \rev{221} & \rev{19,765} \\
\hline
\multirow{6}{*}{$\dfrac{\#Edge}{\#Input}$}
 & \app & \rev{2.183} & \rev{1.635} & \rev{2.538} & \rev{1.429} & \rev{1.652} & \rev{1.160} & \rev{1.622} & \rev{0.733} & \rev{1.147} & \rev{0.471} & \rev{1.569} \\
 & EvoSuite-100s & 0.572 & 0.359 & 0.670 & 0.414 & 0.155 & 0.265 & 0.268 & 0.181 & 0.372 & 0.104 & 0.388 \\
 & EvoSuite-150s & 0.399 & 0.256 & 0.544 & 0.304 & 0.099 & 0.175 & 0.201 & 0.120 & 0.258 & 0.070 & 0.274 \\
 & EvoSuite-200s & 0.308 & 0.210 & 0.418 & 0.251 & 0.085 & 0.149 & 0.154 & 0.080 & 0.192 & 0.059 & 0.213 \\
 & LLM-baseline & \rev{3.204} & \rev{2.383} & \rev{4.634} & \rev{2.1962} & \rev{2.754} & \rev{1.784} & \rev{3.179} & \rev{1.530} & \rev{2.149} & \rev{0.715} & \rev{2.533} \\
 & \rev{\app-CG} & \rev{1.563} & \rev{1.863} & \rev{2.094} & \rev{0.957} & \rev{1.490} & \rev{1.050} & \rev{2.190} & \rev{0.410} & \rev{0.959} & \rev{0.358} & \rev{1.223} \\
\hline
\multirow{6}{*}{\#$Time$}
 & \app & \rev{19,915} & \rev{6,799} & \rev{4,667} & \rev{5,399} & \rev{6,694} & \rev{3,724} & \rev{2,360} & \rev{7,234} & \rev{13,293} & \rev{3,344} & \rev{73,429} \\
 & EvoSuite-100s & 54,500 & 16,900 & 13,100 & 19,500 & 18,500 & 10,600 & 7,000 & 20,100 & 50,500 & 9,800 & 220,500 \\
 & EvoSuite-150s & 81,750 & 25,350 & 19,650 & 29,250 & 27,750 & 15,900 & 10,500 & 30,150 & 75,750 & 14,700 & 330,750 \\
 & EvoSuite-200s & 109,000 & 33,800 & 26,200 & 39,000 & 37,000 & 21,200 & 14,000 & 40,200 & 101,000 & 19,600 & 441,000 \\
 & LLM-baseline & \rev{5,565} & \rev{2,214} & \rev{1,431} & \rev{1,919} & \rev{2,379} & \rev{1,809} & \rev{1,148} & \rev{3,566} & \rev{10,520} & \rev{1,409} & \rev{31,960} \\
 & \rev{\app-CG} & \rev{20,703} & \rev{6,621} & \rev{4,280} & \rev{7,847} & \rev{6,264} & \rev{3,626} & \rev{2,059} & \rev{7,906} & \rev{13,356} & \rev{4,192} & \rev{76,854} \\
\hline
\multirow{6}{*}{$\dfrac{\#Time}{\#Input}$}
 & \app & \rev{5.84} & \rev{9.80} & \rev{6.05} & \rev{4.53} & \rev{7.85} & \rev{7.37} & \rev{5.91} & \rev{7.00} & \rev{4.72} & \rev{6.65} & \rev{6.03} \\
 & EvoSuite-100s & 5.49 & 5.82 & 5.59 & 5.42 & 6.03 & 5.97 & 5.01 & 5.40 & 5.73 & 5.55 & 5.61 \\
 & EvoSuite-150s & 5.70 & 6.10 & 6.09 & 5.66 & 5.97 & 5.80 & 5.98 & 5.79 & 5.81 & 5.78 & 5.82 \\
 & EvoSuite-200s & 5.85 & 6.35 & 5.97 & 5.91 & 5.99 & 6.08 & 6.12 & 6.01 & 5.97 & 5.96 & 5.98 \\
 & LLM-baseline & \rev{4.19} & \rev{13.67} & \rev{11.63} & \rev{4.05} & \rev{7.90} & \rev{9.52} & \rev{17.13} & \rev{12.78} & \rev{9.68} & \rev{10.83} & \rev{7.72} \\
 & \rev{\app-CG} & \rev{4.06} & \rev{9.99} & \rev{4.80} & \rev{4.33} & \rev{6.91} & \rev{5.81} & \rev{6.02} & \rev{4.28} & \rev{3.98} & \rev{6.78} & \rev{4.76} \\
\bottomrule
\end{tabular}
\label{tbl:cov-eff}
\end{table*}

We have obtained \apinum API methods, employing the following strategies for method selection to improve the quality of our datasets.

\begin{enumerate}[leftmargin=10pt]
    \item Exclude methods within abstract classes or interfaces, since the classes or interfaces cannot be instantiated directly.
    \item Exclude methods that only have one basic block, since 100\% edge coverage is guaranteed and meaningless.
    \item Exclude methods inherited from \inlinecode{class Object} (\eg \inlinecode{equals}, \inlinecode{toString}, \inlinecode{hashCode}).
\end{enumerate}

\textit{Implementation}. To demonstrate the feasibility of \app, we have implemented it in Java. Specifically, we utilize \textit{JDT}~\cite{jdt} and \textit{Soot}~\cite{soot} to obtain AST, class hierarchy and call graph.
We employ \textit{langchain}~\cite{Chase_LangChain_2022} to interact with LLMs. It is important to note that while the implementation is specific to Java, the underlying concept of \app can be applied to \cOOPlangs and automated testing frameworks in a more general scene.

\textit{Baselines \& Variant}. We have chosen two baselines, in order to better conduct various experiments.

\begin{itemize}[leftmargin=10pt]
    \item Search-based baseline. EvoSuite~\cite{2011evosuite}, a state-of-the-art tool in the field of SBST, \rev{is still actively maintained and has continuously been incorporating new SBST optimization algorithms since its release. It is widely used in both academia and industry. We choose the latest version released in 2021, Evosuite-v1.2.0, and refer to the time budget used in previous studies}.~\cite{fse2021lingraph}. 
    \item LLM-based baseline. We only provide the signature and code of the library \mut and expect the LLM to output the same format of results of \app directly.
    \item \app-CG (\app with Call Graph). A variant that first obtains the call graph of the target library, and then includes the source code of those methods called within the \mut when constructing the prompt, in order to investigate the impact of ``the source code of the called method'' as mentioned in \Cref{sec:approach}. We include the source code of only one layer of methods invoked by the \mut. \app never exceed the token limits during our evaluation, even though there are no prompt trimming or compression tricks in \app. The prompt of \app-CG is available at our site~\cite{site}.
    \item Symbolic-based baseline. We have tried SPF~\cite{2019svcompspf} with lazy initialization, whose performance and scalability are among the best. We initially run SPF on 35 APIs, but only 4 are able to run. The remaining 31 APIs suffer from issues such as path explosion, insufficient support for collections, arrays and interfaces, etc. As a result, we abandon the comparison with symbolic execution tools.
\end{itemize}

\rev{\textit{Metrics}. We evaluate \app and baselines based on branch coverage and the number of found exceptions. Specifically, we adopt the branch coverage collection module used in EvoSuite to record coverage during each execution. For exception detection, we employ the same module in JQF~\cite{issta2019jqf} to record detected exceptions.}

\begin{figure}[htb]
  \centering
  \includegraphics[width=\columnwidth]{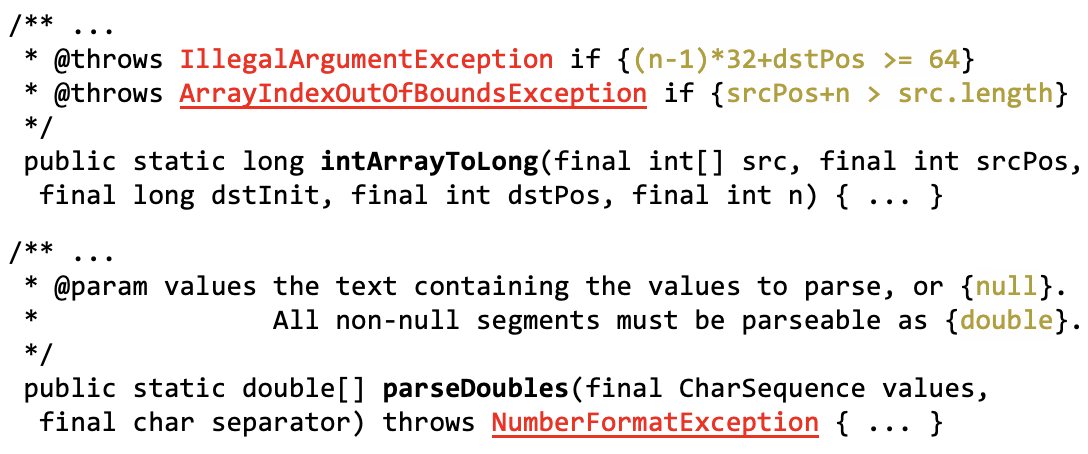}
  \caption{False Positive Exception Examples (Javadoc \& Signature)}
  \label{fig:exception-eg}
\end{figure}

\rev{\textit{Identifying False Positives.} Unlike system testing, API testing may generate false positives. These false positives occur when the generated inputs violate the assumptions of the APIs, leading to exceptions. The API assumptions are typically specified in Javadoc comments. For instance, as shown in ~\Cref{fig:exception-eg}, API \inlinecode{intArrayToLong} from library commons-lang3 assumes their parameters need to meet \inlinecode{srcPos + nInts > src.length} constraint. During testing, inputs that do not meet such constraints can be generated, resulting in false positives. }

\rev{
We identify such false positives based on a convention used in the Java API specification. Specifically, when an API assumption is violated, the type of exception thrown is often specified in the Javadoc comments or the signature of the API.
}

\begin{enumerate}[leftmargin=10pt]
    \item \rev{Javadoc Exceptions. As shown in \Cref{fig:exception-eg}, exception \inlinecode{ArrayIndex}\inlinecode{OutOfBounds}\inlinecode{Exception} for constraint \inlinecode{srcPos+n > src.length} is specified in the Javadoc comments.}
    \item \rev{Signature Exceptions. As shown in \Cref{fig:exception-eg}, exception \inlinecode{Number}\inlinecode{Format}\inlinecode{Exception} is specified in its signature.}
\end{enumerate}

\rev{In our experiments, we take a conservative approach by filtering out all exceptions related to API assumption violations during result reporting. Specifically, for each \mut, we use Soot~\cite{soot} to extract exceptions declared in Javadoc comments and signatures, and exclude these exceptions from the collected data during testing. The related code can be found in our artifact~\cite{site}.}

\textit{Evaluation Environment}. Our experiments run on a 64-bit Linux machine (Ubuntu-22.04) with a 1.8GHz 8-Core AMD Ryzen 7 5700U CPU and 16GB RAM and use an OpenAI API key with 500 RPM to run all experiments. \rev{We use \textit{gpt-3.5-turbo} with a token limit of 16K. To make the output more consistent, we set the temperature to 0.}

\subsection{RQ1: Code Coverage} \label{sec:eval-rq1}
Test inputs with higher code coverage are usually indicative of more comprehensive execution across the API functionalities. In addition, it is critical to generate high-quality inputs stably for API testing. In this study, we evaluate \app from three dimensions,
(1) average code coverage (for individual libraries and overall average),
(2) quality (indicated by the coverage improvement caused by inputs), and
(3) efficiency (indicated by the speed of generating valid inputs).

\rev{\Cref{tbl:cov-eff} presents the results of our experiment:
(1) \#Input.
(i) For the \app and LLM-baseline rows, each number represents the total number of valid inputs generated by the experiment. These inputs are generated by the tool for each API in the corresponding Java library executed once.
(ii) For the EvoSuite-Xs rows, each number represents the total number of inputs generated by running each API for X seconds using EvoSuite.
(2) \#Edge. In \Cref{sec:eval-setup}, for the \inlinecode{\app} and \inlinecode{LLM-baseline} rows, each number represents the total number of distinct edges covered by running each API just once in the corresponding library, which is also identical to the numerator of the ``average code coverage'' metric.
(3) \#Time. Each number represents the total time to generate inputs for all APIs in the corresponding Java library.}

\begin{figure}[tb]
  \centering
  \includegraphics[width=\columnwidth]{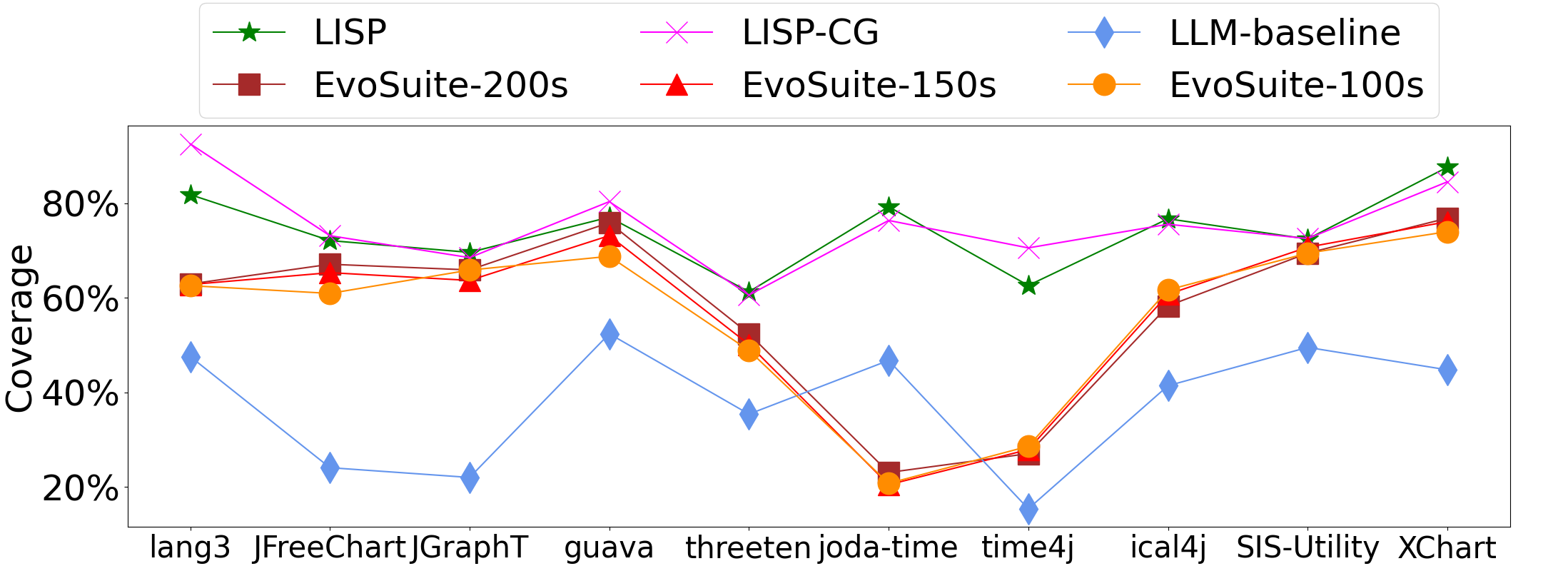}
  \caption{Average code coverage of the \libnum selected libraries.}
  \label{fig:cov-res}
\end{figure}

\begin{figure}[tb]
  \centering
  \includegraphics[width=\columnwidth]{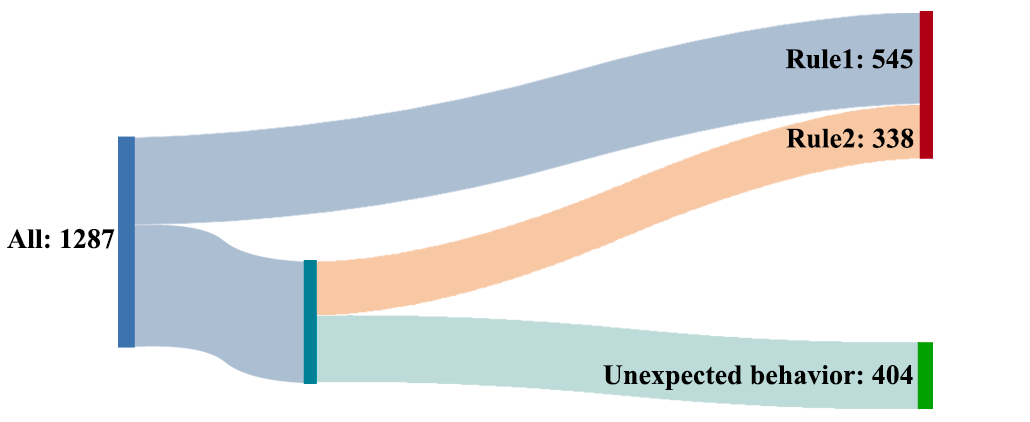}
  \caption{\rev{A sankey diagram illustrating the filtering process that utilizes ``Javadoc Exceptions'' (Rule 1) and ``Signature Exceptions'' (Rule 2) to handle all captured exceptions and errors.}}
  \label{fig:ub-rules}
\end{figure}

\rev{\textit{Result \& Analysis}.
(1) \rev{Average code coverage}. As shown in \Cref{fig:cov-res} and \Cref{tbl:cov-eff}, in the selected \libnum libraries, \app overall outperforms both baselines, with an average code coverage that is \rev{1.25} times that of EvoSuite-100s, \rev{1.22} times that of EvoSuite-150s, and \rev{1.21} times that of EvoSuite-200s. \app-CG achieves similar code coverage with much more inputs.
In addition, as shown in \Cref{tbl:cov-cost}, \app-CG consumes much more tokens.
(2) \rev{Quality}. As shown in \Cref{tbl:cov-eff}, in terms of coverage improvement caused by inputs, \app still overall outperforms both baselines, with fewer inputs but highest edge coverage, whose code coverage improvement per input is \rev{4.04} times that of EvoSuite-100s, \rev{5.73} times that of EvoSuite-150s, and \rev{7.37} times that of EvoSuite-200s. LLM-based baseline generates the smallest number of inputs and attained less edges than \app.
\rev{(3) Efficiency. As shown in \Cref{tbl:cov-eff}, considering the time efficiency for generating valid inputs, EvoSuite outperforms \app because EvoSuite uses search-based algorithms, making it easier to generate valid inputs. However, the efficiency of LLM-based variants is not unacceptable.}}

\textit{Summaries}.
(1) In terms of code coverage, \app outperforms both search-based baseline (\ie EvoSuite) and LLM-based baseline;
(2) In terms of quality, \app outperforms both baselines and achieves the highest coverage improvement per input.
\rev{(3) In terms of efficiency, \app outperforms all EvoSuite variants, but LLM-based approach exhibited less time cost, because \app requires more interaction with LLMs.}

\subsection{RQ2: Usefulness}
The ability to find software vulnerabilities is one of the most effective criterion for judging an automated testing tool.
In this study, we mainly concern three aspects,
(1) statistics (indicated by categories and count of \ubs),
(2) differences (indicated by the diversity in \ubs triggered by \app and baselines), and
(3) vulnerabilities (indicated by findings).

\begin{figure}[tb]
  \centering
  \includegraphics[width=0.95\columnwidth]{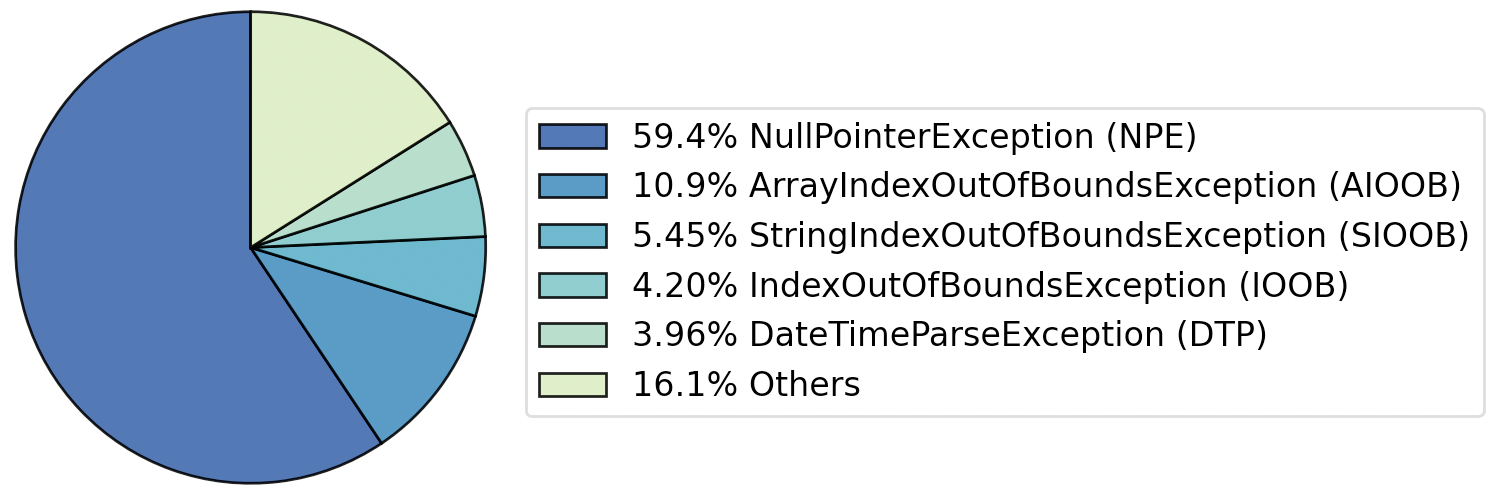}
  \caption{Top 5 \ub types found by \app.}
  \label{fig:ub-type}
\end{figure}

\begin{figure}[tb]
  \centering
  \includegraphics[width=0.95\columnwidth]{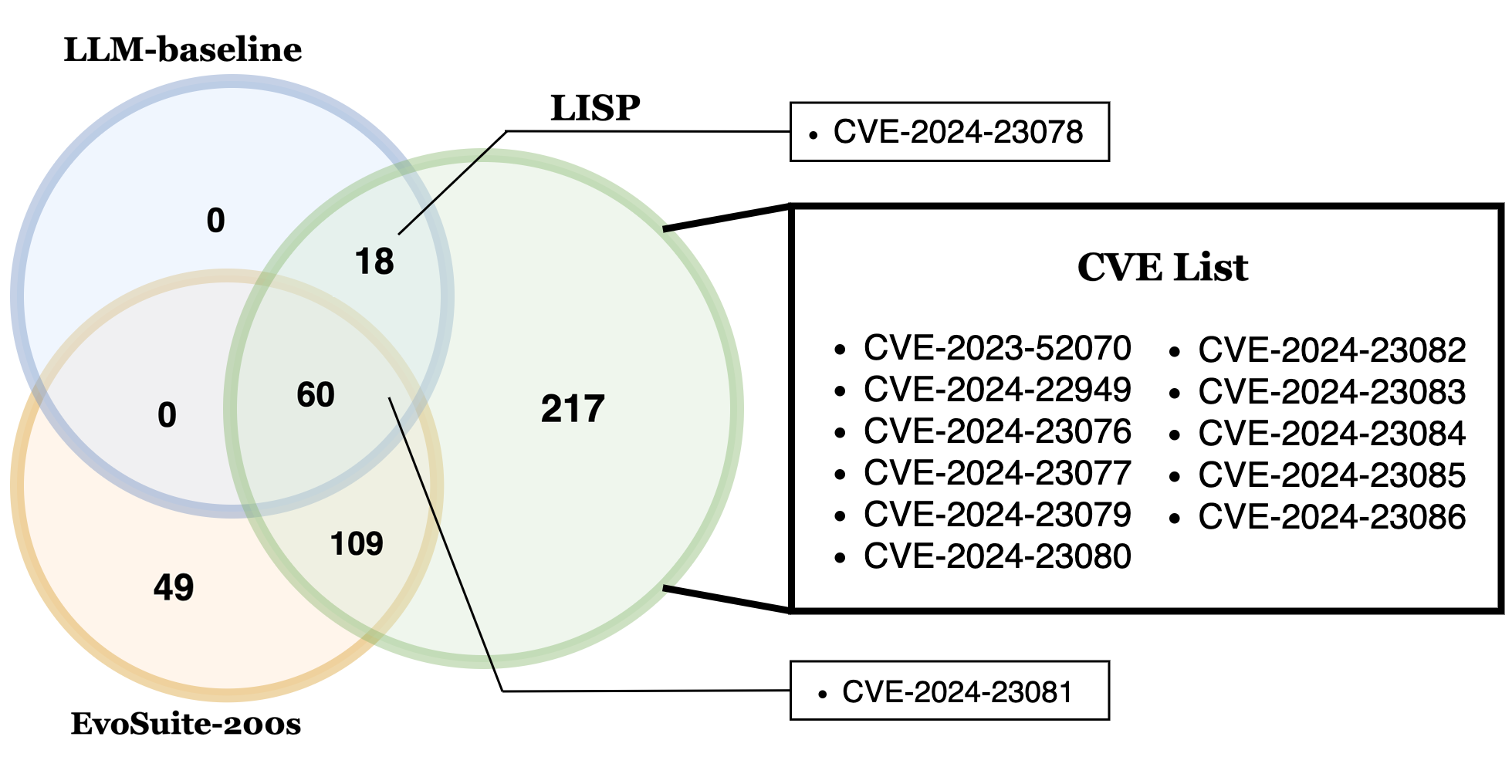}
  \caption{A Venn diagram representing the distribution of \ubs found by \app, \textit{EvoSuite-200s}, LLM-baseline as well as the details of CVEs}
  \label{fig:ub-venn}
\end{figure}

\textit{Result}.
(1) Statistics. \rev{\app captures a total of \ubtotal Java exceptions or errors. As shown in \Cref{fig:ub-rules}, after applying the two filtering rules outlined in \Cref{sec:eval-setup} wherein Rule 1 (``Javadoc Exceptions'') filters out 545 of those and Rule 2 (``Signature Exceptions'') subsequently filters out an additional 338, we finally obtain \ubnum \ubs, totaling \ubtype types.}
As shown in \Cref{fig:ub-type}, it is important to note that ``NPE'' accounts for the largest proportion, reaching 59.4\%, followed by ``index out of bound'' is the second most common, accounting for a total of 35.75\% (``AIOOB'' (10.9\%) + ``SIOOB'' (5.45\%) + ``IOOB'' (4.20\%)). 
We also record the \ubs captured by baselines, among which LLM-baseline captures 9 types of \ubs, a total of 78, while EvoSuite with a time budget of 200s (the best in EvoSuite) captures 13 types of \ubs, a total of 217 (but due to space constraints we no longer tabulate). 
(2) Differences. As depicted in \Cref{fig:ub-venn}, \app captures all the \ubs identified by LLM-baseline, which is expected given that \app has extended the capabilities of LLM-baseline through prompt-engineering. In absolute terms, \app only misses 49 \ubs that EvoSuite detects, while EvoSuite fails to identify 235 \ubs that \app detects.
(3) Vulnerabilities. We conduct a case-by-case study on \ubs found during our evaluation. Then, we identify vulnerabilities and report them. As shown in \Cref{fig:ub-venn}, \cvenum \rev{previously undiscovered} vulnerabilities are identified as CVEs to date, all of which are detectable by \app, with 11 of those being unique findings of our approach. \Cref{tbl:cve-detail} shows that ``NPE'' still accounts for the largest proportion in identified CVEs. In addition, a ``StackOverflowError'' is identified as a CVE. 
\begin{table}[htb]
\caption{The type and ID of found CVEs.}
\centering
\fontsize{7}{10}\selectfont 
\begin{tabular}{ll}
\toprule
Categories & CVE-ID \\
\midrule
NullPointerException &
\begin{tabular}[c]{@{}l@{}}
    CVE-2024-22949 CVE-2024-23076 \\
    CVE-2024-23078 CVE-2024-23080 \\
    CVE-2024-23081 CVE-2024-23083 \\
    CVE-2024-23085 \\
\end{tabular} \\
\midrule
ArrayIndexOutOfBoundsException &
\begin{tabular}[c]{@{}l@{}}
    CVE-2023-52070 CVE-2024-23077 \\
    CVE-2024-23079 CVE-2024-23084 \\
\end{tabular} \\
\midrule
StringIndexOutOfBoundsException & CVE-2024-23082 \\
\midrule
StackOverflowError & CVE-2024-23086 \\
\bottomrule
\end{tabular}
\label{tbl:cve-detail}
\end{table}

\begin{table*}[htb]
\caption{\rev{Details of Parsing Failure, Compiling Failure and Token Consumption in RQ3.}}
\centering
\fontsize{6}{10}\selectfont 
\begin{tabular}{c|c|cccccccccc|c}
\toprule
\multirow{2}{*}{Metrics} & \multirow{2}{*}{Indicators} & \multicolumn{10}{c|}{Libraries} & \multirow{2}{*}{Overall} \\
\cline{3-12}
 &  & commons-lang3 & JFreeChart & JGraphT & guava & joda-time & threeten & time4j & iCal4j & SIS-Utility & XChart &  \\
\midrule
\multirow{3}{*}{\shortstack[c]{$\#API$ \\ $Failed$}}
& \rev{\app} & \rev{2 / 545} & \rev{4 / 195} & \rev{1 / 169} & \rev{1 / 131} & \rev{5 / 185} & \rev{5 / 106} & \rev{9 / 70} & \rev{0 / 201} & \rev{2 / 505} & \rev{0 / 98} & \rev{29 / \apinum} \\
& \rev{\app-CG} & \rev{3 / 545} & \rev{7 / 195} & \rev{6 / 169} & \rev{2 / 131} & \rev{7 / 185} & \rev{7 / 106} & \rev{4 / 70} & \rev{0 / 201} & \rev{3 / 505} & \rev{0 / 98} & \rev{39 / \apinum} \\
& \rev{LLM-baseline} & \rev{4 / 545} & \rev{0 / 195} & \rev{0 / 169} & \rev{0 / 131} & \rev{1 / 185} & \rev{0 / 106} & \rev{1 / 70} & \rev{0 / 201} & \rev{3 / 505} & \rev{0 / 98} & \rev{9 / \apinum} \\
\hline
\multirow{3}{*}{\shortstack[c]{$\#Invalid$ \\ $Input$ (\%)}}
& \rev{\app} & \rev{42.04\%} & \rev{62.57\%} & \rev{57.21\%} & \rev{39.11\%} & \rev{49.91\%} & \rev{43.45\%} & \rev{40.27\%} & \rev{59.74\%} & \rev{50.96\%} & \rev{57.98\%} & \rev{49.82\%} \\
& \rev{\app-CG} & \rev{31.52\%} & \rev{51.32\%} & \rev{35.13\%} & \rev{32.45\%} & \rev{36.60\%} & \rev{41.08\%} & \rev{33.72\%} & \rev{35.41\%} & \rev{42.02\%} & \rev{45.69\%} & \rev{37.02\%} \\
& \rev{LLM-baseline} & \rev{38.12\%} & \rev{81.05\%} & \rev{78.15\%} & \rev{29.40\%} & \rev{55.93\%} & \rev{61.54\%} & \rev{80.41\%} & \rev{50.18\%} & \rev{34.44\%} & \rev{49.42\%} & \rev{49.68\%} \\
\hline
\multirow{3}{*}{\shortstack[c]{$\#Token$ \\ $Input$}}
& \rev{\app} & \rev{19,877,517} & \rev{3,020,835} & \rev{2,341,596} & \rev{3,485,580} & \rev{3,306,834} & \rev{4,247,010} & \rev{2,804,631} & \rev{8,053,296} & \rev{15,889,005} & \rev{3,926,481} & \rev{66,952,785} \\
& \rev{\app-CG} & \rev{20,764,098} & \rev{5,544,456} & \rev{4,297,776} & \rev{6,397,452} & \rev{6,069,378} & \rev{5,219,025} & \rev{3,446,526} & \rev{9,896,454} & \rev{17,849,982} & \rev{4,825,137} & \rev{84,310,284} \\
& \rev{LLM-baseline} & \rev{2,931,549} & \rev{909,048} & \rev{704,646} & \rev{1,048,902} & \rev{995,112} & \rev{570,174} & \rev{376,530} & \rev{1,964,694} & \rev{4,936,167} & \rev{957,909} & \rev{15,394,731} \\
\hline
\multirow{3}{*}{\shortstack[c]{$\#Token$ \\ $Output$}}
& \rev{\app} & \rev{1,726,830} & \rev{324,240} & \rev{251,334} & \rev{374,121} & \rev{354,936} & \rev{435,843} & \rev{287,820} & \rev{826,458} & \rev{1,255,674} & \rev{402,951} & \rev{6,240,207} \\
& \rev{\app-CG} & \rev{2,127,645} & \rev{411,198} & \rev{318,741} & \rev{474,462} & \rev{450,129} & \rev{490,041} & \rev{323,613} & \rev{929,232} & \rev{1,751,994} & \rev{453,057} & \rev{7,730,112} \\
& \rev{LLM-baseline} & \rev{540,330} & \rev{167,553} & \rev{129,879} & \rev{193,329} & \rev{183,414} & \rev{105,093} & \rev{69,399} & \rev{353,664} & \rev{888,558} & \rev{172,434} & \rev{2,803,653} \\
\hline
\multirow{3}{*}{\#$Cost$}
& \rev{\app} & \rev{12.529812} & \rev{1.9969238} & \rev{1.5479114} & \rev{2.3041429} & \rev{2.1859817} & \rev{2.7774227} & \rev{1.8341470} & \rev{5.2666223} & \rev{9.828611} & \rev{2.5678059} & \rev{42.8393807} \\
& \rev{\app-CG} & \rev{13.574352} & \rev{3.389278} & \rev{2.627192} & \rev{3.9107060} & \rev{3.710157} & \rev{3.344790} & \rev{2.208823} & \rev{6.3424792} & \rev{11.553664} & \rev{3.092353} & \rev{53.7537942} \\
& \rev{LLM-baseline} & \rev{2.2764086} & \rev{0.705896} & \rev{0.547173} & \rev{0.814495} & \rev{0.772726} & \rev{0.442751} & \rev{0.292383} & \rev{1.5129235} & \rev{3.801126} & \rev{0.737644} & \rev{11.9035261} \\
\bottomrule
\end{tabular}
\label{tbl:cov-cost}
\end{table*}

\textit{Case Study: CVE}. To better illustrate the role of \app in triggering \ubs and discovering vulnerabilities, we select one of CVEs that \app found during our experiments (CVE-2024-23086). As shown in \Cref{lst:cve-reason}, \inlinecode{modPow} is an instance method of \inlinecode{DoubleModMath}, used to calculate the result of ``$a^n \mod m$'', where $m$ denotes the return value of \inlinecode{getModulus}. When $n = 0$, it naturally returns $1$ directly. When $n < 0$, due to \textit{Fermat's little theorem}~\cite{enwiki-fermat}, $a^m \equiv 1 (\mod m)$ holds when $m$ is prime, therefore $a^{m-1+n} = a^n (\mod m)$. In \inlinecode{modPow}, it employs recursive calls to gradually transform $n$ into $m-1+n$, until $l*(m-1)+n > 0$, where $l$ denotes the number of recursive layers. For power calculation, it is not wrong in mathematics. However, in the field of programming, the size of the stack is limited. If $m-1$ is excessively small and $n$ is a negative number with an extremely large absolute value (\eg $p = 2, n = -100,000.0$), too many recursive calls will lead to a stack overflow.

\begin{lstlisting}[caption = {CVE-2024-23086: StackOverflowError due to recursive calls}, label = {lst:cve-reason}]
// DoubleModMath.java
public final double modPow(double a, double n) {
    if (n ==0) { return 1; }
    else if (n < 0) {
        return modPow(a, getModulus() - 1 + n);
    }
    // ignore some code
    return r;
}

// DoubleElementaryModMath.java
public final double getModulus() {
    return this.modulus;
}
private double modulus;
\end{lstlisting}

\begin{lstlisting}[caption = {POC of CVE-2024-23086}, label = {lst:cve-poc}]
public class TestModPow {
    @Test
    public void testModPow() {
        DoubleModMath dmm = new DoubleModMath();
        // assign "2" to modulus
        dmm.setModulus(2);
        // throw java.lang.StackOverflowError
        dmm.modPow(4, -1000000.0);
    }
}
\end{lstlisting}

We attribute the generation of this input originates to LLMs' understanding of code and real-world knowledge. At the \impllevel, LLMs recognize the significance of recursive calls when $n < 0$. At the conceptual-level, LLMs consider that when $m$ is small, \inlinecode{modPow} demands a substantial number of recursive calls to enter the subsequent logic, by combining real-world knowledge from \textit{Fermat's Little Theorem} with the possible reasons of stack overflow.

\rev{\textit{Case Study: LISP‘s miss.}}
\rev{To better illustrate the limitations of \app and explore how to further enhance the current \app capabilities, we conduct a case-by-case analysis on the 49 \ubs that EvoSuite can trigger but \app misses.
Finally, we find that ``index out of bound'' accounts for 53.0\%, while ``NPE'' accounts for 22.4\%.}

\begin{lstlisting}[caption = {\rev{An \ub that EvoSuite detected but \app failed}}, label = {lst:evosuite-better}]
// Strings.java
public static String toString(
        final Class<?> classe,
        final Object... properties) {
    final StringBuilder buffer =
        new StringBuilder(32)
            .append(Classes.getShortName(classe))
            .append('[');
    // ignore some code
    for (int i=0; i<properties.length; i++) {
        final Object value = properties[++i];
        if (value != null) {
            // ignore some code
        }
    }
    return buffer.append(']').toString();
}
\end{lstlisting}

\rev{As shown in \Cref{lst:evosuite-better}, \inlinecode{toString} is a static method that EvoSuite has successfully triggered an exception for, but LISP misses. This method takes a \inlinecode{Class} object and a ``varargs'' of \inlinecode{Object} as parameters.
By reviewing the code, we find that if the number of arguments passed to the \inlinecode{properties} is not even, an \inlinecode{ArrayIndexOutOfBoundsException} will be triggered at Line 11. We have summarized two reasons why \app fails to trigger this exception.
(1) The current prompt design of \app lacks special treatment for arrays, resulting in not good enough performance in detecting ``index-out-of-bound'' exceptions.
(2) \app generates fewer inputs and undergoes a certain randomness.
In the future, we will further enhance \app in these aspects.}

\textit{Summaries}.
(1) In terms of statistics, \app triggers \ubnum \ubs, a total of 18 types. In addition, \app fully covers the \ubs triggered by LLM-baseline, while also triggering 77.5\% of the \ubs triggered by search-based baseline (\ie EvoSuite-200s). In comparison, the search-based baseline only triggered 41.8\% of the \ubs.
(2) In terms of vulnerabilities, \app identifies \cvenum \rev{previously undiscovered} CVEs in total, and 11 of them are derived from the \ubs that both baselines fail to trigger. 

\subsection{\rev{RQ3: Cost}}
\rev{Two main aspects of cost need to be considered when using LLMs to generate inputs.
(1) Failures. How many APIs the LLM cannot correctly provide parsable answers due to hallucinations. Additionally, how much of the generated code is actually not executable.
(2) Token consumption. Whether the token cost of interacting with the LLM is within an acceptable range.}

\Cref{tbl:cov-cost} presents the results of experiments.
(1) \#API Failed. Each ratio represents ``the number of APIs that failed to generate any inputs'' / ``the total number of APIs''.
(2) \#Invalid Input. Each number represents the ratio to the total number of inputs generated cannot be run directly.
(3) \#Token Input and \#Token Output. Each number represent the amount of tokens consumed.
(4) \#Cost. It is obtained according to ``\#Token Input'' and ``\#Token Output'', which is based on the OpenAI billing standard~\cite{openapipricing} in US dollars. 

\rev{\textit{Result}.
(1) Failures. As shown in \Cref{tbl:cov-cost}, all the LLM-based variants demonstrate stable output under \textit{gpt-3.5-turbo}.
However, nearly 50\% inputs generated by both \app and the baseline cannot be run directly, while \app-CG has around 10\% lower failure rate.
(2) Token consumption. The pricing of \textit{gpt-3.5-turbo} we use is ``US\$0.50 per 1M input tokens'' and ``US\$1.50 / 1M output tokens''~\cite{openapipricing}. As shown in \Cref{tbl:cov-cost}, after testing \apinum APIs, \app incurs a cost of \$42.84, with 66.95M tokens as input and 6.24M tokens as output. Also, we run \app-CG and LLM-baseline.
\begin{itemize}[leftmargin=10pt]
    \item For \app-CG (\app with deeper functions), on the input side, the token consumption increases significantly, over 80\% on libraries like JFreeChart and guava. The overall input token consumption across the libraries increased by more than 20\%.
    On the output side, the token consumption increases slightly, since the output formats are not changed.
    \item For LLM-baseline, both the input token and output token consumption decrease significantly, because the LLM-baseline refrains from frequently interacting with the LLM on the task of constructor selection.
\end{itemize}
}

\rev{\textit{Analysis}}.
\rev{(1) The \app and \app-CG can yield parsing failures during the whole workflow depicted in \Cref{fig:approach}. In contrast, the LLM-based baseline only outputs the results directly, which reduces interactions and fewer parsing failures.
(2) Actually, nearly 50\% failure rate is acceptable~\cite{alshahwan2024fsellmtestmeta}. The failure rate is stable across \libnum libraries for \app and \app-CG. \app-CG provides the LLM with more context and results in a lower failure rate. However, the baseline exhibits a large variance, possibly due to a lack of task decomposition for input generation.}

\rev{\textit{Summaries}.
(1) In terms of failures, nearly half of the inputs generated by \app are not runnable, but it is still acceptable~\cite{alshahwan2024fsellmtestmeta}.
(2) In terms of token consumption, \app naturally consumes more tokens than the LLM-based baseline, but overall, it still remains within a reasonable range. Furthermore, \app-CG consumes much more token than \app, although it achieves more stable output and higher coverage.
}

\subsection{RQ4: Ablation Study}
In this study, we explore the role of each part within \app and evaluate their contributions to the overall approach. In \Cref{sec:motivation}, we have summarized the input object generation process and the importance of \ecp. 
Here, we design an ablation study that consists of three parts (no ISP+TDA since we need instantiation statements to generate input objects and test drivers). 

\begin{itemize}[leftmargin=10pt]
    \item ISP+OI (without \tdtda), a variant that cannot select the appropriate constructors step by step, and expects LLMs to generate inputs directly.
    \item TDA+OI (without \ecp), a variant that only simulates the process of input generation solely through \tdtda and \buoi.
\end{itemize}

\begin{figure}[tb]
  \centering
  \includegraphics[width=\columnwidth]{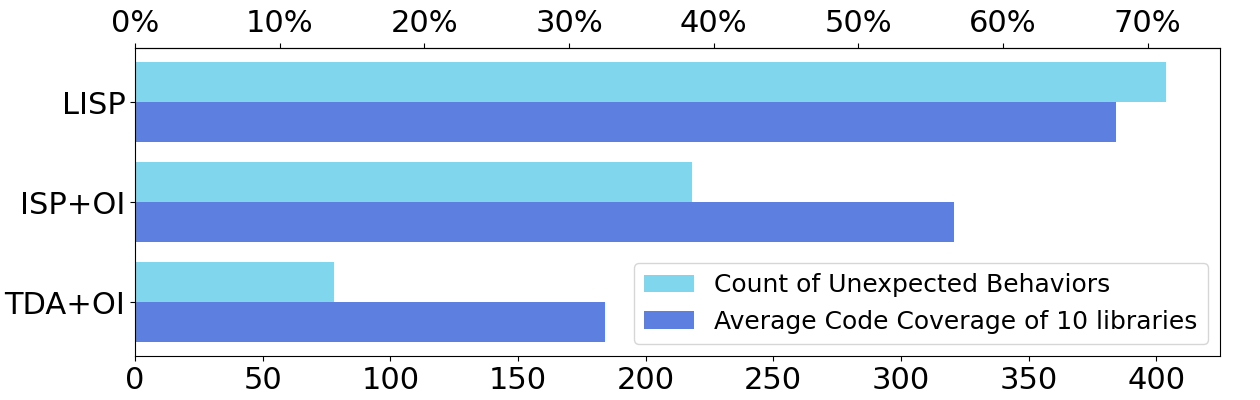}
  \caption{The experiment results of ablation study.}
  \label{fig:ablation}
\end{figure}

\textit{Result \& Analysis}. As shown in \Cref{fig:ablation}, we can see that \app overall performs the best, followed by \wotdaoi and \woisp.
(1) ISP+OI brings a certain decline (code coverage of ISP+OI is 56.60\%, which is 83.46\% of LISP). Based on the results of \ecp, the LLM still generates inputs purposefully. However, the absence of TDA in this variant contributes to the generation of invalid objects.
(2) TDA+OI brings a significant decrease both in ``average code coverage'' and in ``\ubs'' (code coverage of TDA+OI is 32.52\%, which is 47.95\% of LISP). This further indicates that the LLM lacks essential directives in the process of constructor selection, due to the absence of \ecp.

\textit{Summaries}. \Ecp and \tdtda are both effective and contribute significantly to \app.
(1) \Ecp significantly improves the code coverage of the input objects generated by LLMs.
(2) \Tdtda assists LLMs in effectively understanding nested reference types and generating valid objects.

\section{\rev{Limitations}}
\label{sec:limit}

\rev{(1) Interpretability challenges. Since the selected LLM used in \Cref{sec:evaluation} is closed-source, we cannot provide a set of state-of-the-art prompts. }
\rev{(2) Complicated API interactions. Currently, \app cannot generate a sequence of API calls to handle interactions between APIs. This is our future research direction.}
(3) Currently used drivers. The term ``inputs'' of a method should also include environment variables, system configurations, and more. Our drivers can merely generate arguments for the \mut. 
\rev{(4) Document-enhanced prompt engineering. Currently, we only include code comments when feeding API code into LLMs. We believe that it would be beneficial to integrate relevant documentation, and we plan to investigate retrieval-augmented generation (RAG) to achieve such an integration in future work.}
\section{Related Work}
\label{sec:related-work}

\subsection{Input Generation}
Input object generation is a crucial component of automatic test-suite generation that has received significant attention from researchers. Over time, various techniques have been employed in the field of object-oriented input generation~\cite{2011evosuite, 2018icsesushi, fse2021lingraph, ARCURI20083075, 2015igoop}. Gordon Fraser \etal developed \textit{EvoSuite}~\cite{2011evosuite}, which is considered the state-of-the-art SBST tool. To further improve performance, other research projects promote the search-based approaches through advanced algorithms.~\cite{2010sbstsurvey, 2021sbstchallenge}. For instance,
Yun Lin \etal developed \textit{EvoObj}~\cite{fse2021lingraph} that constructs an ``object construction graph'' via static analysis to generate a test seed template. Harrison Green \etal developed \textit{GraphFuzz}~\cite{icse2022graphfuzz} that mutates the ``dataflow graph'' to generate more test templates and unit tests.

\subsection{\LLMsfn}
Present \LLMsfn\ (LLMs) are typically developed through a two-step process~\cite{openai2023gpt4}. Initially, they are trained on massive quantities of diverse text data, enabling them to capture the intricacies of language and acquire a wide range of knowledge~\cite{Radford2019LanguageMA, chowdhery2022palm, gao2020pile, austin2021program, nijkamp2023codegen}. Subsequently, these pre-trained models undergo a fine-tuning phase using additional datasets, further refining their understanding and text generation abilities, allowing them to possess extensive knowledge, language understanding and text generation capabilities~\cite{wei2022finetuned, chung2022scalingfinetune, touvron2023llama}. In addition, they possess the capabilities to generate consistent and appropriate text results for various natural language processing tasks, such as text generation~\cite{nijkamp2023codegen, jiao2023chatgpttranslation}, information extraction~\cite{ma2023largeextraction}, etc.

Recently, LLMs are applied to various fields of secure software development life-cycle, including implementation~\cite{chen2021codex, fried2023incoder}, maintenance~\cite{prenner2021automatic, xia2023automated} and testing~\cite{lemieux2023codamosa, deng2023large, deng2023fuzzgpt}. To interact with LLMs more efficiently~\cite{brown2020language, reynolds2021prompt}, prompts with task definitions and demonstrations are typically employed for better performance~\cite{touvron2023llama, wei2022finetuned, wei2023chainofthought}. In the context of software testing, LLMs are often provided with zero-shot or few-shot prompts to synthesize input generators, method invocations and assertions~\cite{deng2023large, brown2020language}.

\section{Conclusion and Future Work}
\label{sec:conclusion}

In this paper, we explore the potential of LLMs in the field of \ecp testing. Compared to the existing techniques, we have utilized the information in the code, which plays a crucial role in constructing high-quality inputs. Our experiments show that our approach achieves higher coverage, higher efficiency and stronger ability to find vulnerabilities.

In future work, we aim to build upon and extend these findings. We plan to address the sophisticated challenges of API testing, such as test generation involving multiple APIs and API testing in microservices.
Furthermore, we will delve into the combination between software testing and LLM-related emerging technologies (\eg Agents), in order to explore the boundary of automated software testing. We hope that LLMs can enable the automated design and execution of test cases, the comprehensive analysis of results, and even the suggestion of improvements.
In this manner, we will refine not only the efficacy and accuracy of automated tests but also their scalability across diverse and complex software systems.

\section*{Acknowledgements}
We thank the anonymous reviewers for their insightful comments and suggestions. We also thank Yannic Noller for his valuable discussion and feedback on the symbolic execution tool SPF. This work was supported by National Key R\&D Program of China (2023YFB4503805). 
\clearpage


\balance



\end{document}